\documentclass[12pt]{article}
\newcommand{\be}{\begin{equation}}
\newcommand{\ee}{\end{equation}}
\begin{document}
\baselineskip=24 pt
\begin{center}

{\large {\bf The New Planck Scale: Quantized Spin and Charge Coupled to 
Gravity}} 

This essay received an "honourable mention" in the 2003 Essay Competition of the 
Gravity Research Foundation

\end{center}

\vskip1.5truecm 

\begin{center}
  F. I. Cooperstock 
 \\{\small \it Department of Physics and Astronomy, University 
of Victoria} \\
{\small \it P.O. Box 3055, Victoria, B.C. V8W 3P6 (Canada)}\\
V. Faraoni
\\

{\small \it Physics Department, University of Northern British
Columbia\\
3333 University Way, Prince George, B.C. V2N 4Z9 (Canada)}\\

{\small \it e-mail addresses: cooperstock@phys.uvic.ca, vfaraoni@unbc.ca}
\end{center} 
\begin{abstract} 
In the standard approach to defining a Planck scale where gravity is brought 
into the quantum domain, the 
Schwarzschild gravitational radius is set equal to the Compton wavelength. 
However, ignored thereby are the charge and spin, the fundamental quantized 
aspects of matter. The gravitational and null-surface radii of the Kerr-Newman 
metric are used to introduce spin and charge into a new extended Planck scale. 
The 
fine structure constant appears in the extended Planck mass and the recent 
discovery of the $\alpha$ variation with the evolution of the universe adds 
further significance. An extended Planck charge and Planck spin are derived.
There is an intriguing suggestion of a connection with the $\alpha$ value 
governing high-energy radiation in Z-boson production and decay.
\end{abstract} 

Traditionally, one derives the Planck mass by equating the gravitational radius 
$2Gm/c^2$ of a Schwarzschild mass with its Compton wavelength $\hbar/mc$. The 
body has no spin and no charge yet spin and charge are the fundamental quantized 
aspects of matter. Since the Planck scale is to reflect the \textit{quantized} 
union of gravity with matter, surely spin and charge should be incorporated. 
Fortunately, we have the important couplings of spin and charge to gravity in 
the form of the Kerr-Newman \cite{KerrNewman} metric. The gravitational radius 
corresponds now to the upper sign in  
\begin{equation} \label{eqp3} 
r_{\pm } = \frac{G}{c^2} \left( m \pm \sqrt{m^2 - \frac{q^2}{G}  -
\frac{c^2}{G^2} \, a^2} \,\, \right) \; 
\end{equation} 
and the second radius with the lower sign is the equally interesting
``null  surface" radius.
To connect with the quantum domain, we quantize
the charge in units of the charge $ e $ of the electron and
the angular momentum  in units of the fundamental quantum of angular
momentum  $\hbar$, with respective quantum numbers $N$
and $s$: 
\begin{equation} \label{eqp4} 
q=N \, e \;, \;\;\;\;\;\;\;  a=s \, \frac{\hbar}{m} \;. 
\end{equation} 
(Note that the $m$ appears again through the spin.) Setting the
Kerr-Newman event horizon and null surface
(eq.~(\ref{eqp3})) 
radii of  the particles equal to their
Compton wavelengths, and substituting the quantized charge and spin from
eq.~(\ref{eqp4}), we have
\begin{equation} \label{eqp4a}
\frac{\hbar}{mc} = \frac{G}{c^2} \left( m \pm \sqrt{m^2
-\frac{N^2 e^2}{G} - \frac{c^2 \hbar^2 \, s^2}{G^2m^2}} \,\, \right) \;.
\end{equation}
Solving for $m$, we find that the mass which we now refer to as the
\textit{extended Planck mass} $m_{plex}$ is
\begin{equation}  \label{eqp5}
m_{plex} = m_{pl} \, \sqrt{ \frac{2(1+s^2)}{2 -\alpha N^2 }} \; ,
\end{equation}
for both cases, where $\alpha \equiv  e^2/\hbar c \simeq 1/137 $ is the fine 
structure constant
and  $m_{pl}$  designates the standard Planck mass $\sqrt{c\hbar/2G}$ .
From the extended Planck mass, the new Planck length and Planck time, i.e. the 
complete new Planck scale is found in the usual manner.

By eq.~(\ref{eqp5}), the presence of either spin or charge leads to
an increase in the value of $m_{plex}$ as compared to the traditional
$m_{pl}$. It is also interesting to find that these two fundamental quantities 
of physics, the (now extended) Planck scale and the fine structure constant, are 
actually linked. Moreover, the presence of the fine structure constant in
eq.~(\ref{eqp5}) 
provides an additional source of interest, given the current focus upon its 
apparent slow variation in time \cite{Webbetal}-\cite{Murphyetal}. 
Following recent claims \cite{Webbetal}-\cite{Murphyetal} that the value
of the fine structure constant underwent changes during the last half of
the history of the universe, we focus on the possibility that
$\alpha$ could have had a considerably different value in the still more
distant past. If $\alpha$ undergoes significant variations, then $m_{plex}$ does 
as well. Although rather
unorthodox in the low-energy regime, this idea appears quite naturally in
the context of renormalization, in which the coupling ``constants'' are
actually running couplings. In the standard model, the early
universe expands and cools precipitously in its very first instants
when it emerges from the big bang, and the energy scale drops substantially, 
allowing for significant variations in the values of the running couplings.

If the fine structure ``constant'' changes at all,
a change in either $c$ or  $e$ could be responsible~-~see
Refs.~\cite{debate} for a debate on the two possibilities.   A
time-varying 
$\alpha$ can be accomodated in the
context of varying speed of light cosmologies, of which
many proposals have appeared recently \cite{Moffat}-\cite{VSL5}.
While the reported variation 
of $\alpha $ over the last $10^{10}$  years is minute (of the order
of $10^{-5}$ \cite{Webbetal}-\cite{Murphyetal}) and the variation of
fundamental constants is restricted by
primordial nucleosynthesis, it is quite conceivable that more radical
changes could have occurred earlier in the history of the universe.
Although the 
current evidence points to a small increase in $\alpha$ as we go forward in time 
over the time scale thus far surveyed, the essential point is that there is 
variation and this variation could have been one of decrease from a larger value 
at a still earlier time. 

To fix our ideas, suppose that $N=5$ and $s$ is of order unity. Then, if
at sometime in the past, $\alpha$ assumed a value close to $ 8 \cdot 10^{-2}$
(approximately one order of magnitude larger
than its present value), the value of the extended Planck mass $m_{plex}$
would have
been many orders of magnitude larger than its present-day value,
regardless of
the value of the quantum number $s$ (larger values of $N$ lead to large
effects for smaller variations of $\alpha$).

Extremal values are generally useful to gain insight and hence 
we note that the critical upper-limit $N$ value in 
eq.~(\ref{eqp5}) 
is $N=16$ for the present $\alpha$ value of 1/137.036. With this $N$ 
value, the 
extended Planck mass becomes infinite for an $\alpha$ value of 1/128. 
Interestingly, the $\alpha$ value governing high-energy radiation in Z-boson 
production and decay has been measured to be 1/127.934. The Z-boson is 
electrically neutral. Could it be that the connection to the \textit{extremal} 
value reflects this neutrality? Recalling the history of theorizing about the 
number 137, we see in this that there 
really may be some connection between fundamental constants and integers.

It is to be noted that the scope for the extension of the Planck scale is 
severely limited if one were to be restricted to the choice of the event
horizon 
radius $r_{+}$ as opposed to the null surface radius
$r_{-}$. From  eq.~(\ref{eqp4a}) with the positive sign in front
of the square root, we find the inequality
\begin{equation} \label{eqp5a}
\frac{\hbar}{ mc} - \frac{Gm}{c^2} \geq 0
\end{equation} 
and hence, with eq.~(\ref{eqp5})
\begin{equation} \label{eqp5b}
m_{pl} \leq m_{plex} \leq \sqrt{2} \, m_{pl}
\end{equation}

These conditions in conjunction with eq.~(\ref{eqp5}) place the following 
restrictions on the allowed spin and charge quanta:

\begin{equation} \label{eqp7}
s^2 + N^2\alpha \leq 1 \;, \;\;\; \; \; \; \; N^2 \alpha < 2,
\end{equation}
and they lead to a spectrum of spin/charge values. The allowed values of $s$ and 
$N$ for $\alpha = 1/137$ are

a) for $s=0$, $ N \leq 11$

b) for $s=1/2$, $N \leq 10$ 

c) for $s=1$, $N=0$.

Spin-two is not allowed in this case which might evoke some 
surprise as the graviton is seen as a spin-two boson. However the extended
Planck 
mass, as the traditional Planck mass, is very large whereas the graviton
mass is zero to a very high level of accuracy ($m_{graviton} <
10^{-59}$~g). These are very different concepts.

Given the new extended approach, it is natural to introduce an extended {\em 
Planck 
charge} and a {\em Planck spin}. These quantities could be defined by assuming 
that the ``Planck
particle'' considered is an extremal black hole, i.e. one defined by
\be
m^2=\frac{q^2}{G} + \frac{c^2}{G^2}\; a^2 
\ee
(corresponding to the equality in (\ref{eqp7})) that is maximally
charged ($s=0$, $q=q_{max}$) or maximally rotating
($q=0$, $s=s_{max}$).  These requirements yield the extended Planck quantities
\be
q_{plex}=\frac{e}{\sqrt{\alpha}}\simeq 11.7 \, e \;, \;\;\;\;\;\;\;\;\;\;\;
s_{plex}=1 
\ee
(corresponding to the Planck angular momentum $L_{plex} = \hbar$ and now 
allowing for non-integral $N$).
While $q_{plex} $ is large but not extraordinarily so, $L_{plex}$ is rather 
ordinary on the scale of particles familiar at an energy much lower than the 
Planck
scale. 

According to the third law of black hole thermodynamics, an extremal black
hole corresponds to zero absolute temperature, and is an unattainable
state. If the third law survives in the Planck regime, the
values of $N$ and $s$ are even further restricted, and the first of
(\ref{eqp7}) should read as a strict inequality.

If one considers instead the null surface of radius $r_{-}$ defined by
eq.~(\ref{eqp3}) and with (\ref{eqp5}), the inequalities
\begin{equation} \label{eqp8}
s^2 + N^2\alpha \geq 1 \;, \;\;\;\;\;\;\;\;\; N^2\alpha <2
\end{equation}
follow.

In this case, the allowed values of $s$ and $N$ for $\alpha = 1/137$ are,

a) for $s=0$, $12 \leq N \leq 16$

b) for $s=1/2$, $11 \leq N \leq 16$

c) for $s=1$, $0 \leq N \leq 16$

d) for $s=2$, $0 \leq N \leq 16$

In this case, spin two is readily allowed and with the extremal $N=16$ value, 
the $\alpha$ value of 1/128 gives an infinite $m_{plex}$

Finally, a comment is in order regarding the frequently mentioned observation 
that the natural length scale of ``grand unification", the merging of the strong 
and the electroweak interactions, is only a few orders of magnitude larger than 
the standard Planck length scale. Thus, the suggestion arises that ultimately, 
gravitation may hold the key to a final ``super-grand unification", the 
unification of all the interactions. It must be remarked that any spin or 
$\alpha$ modification in the new Planck scale can only increase the mass scale 
and hence lower the length scale. The possibility to be faced is that 
gravitation may remain disjoint from the other interactions in nature.

{\small {\bf Acknowledgments:} This work was supported in part by a grant
from the Natural Sciences and Engineering Research Council of Canada.}

\clearpage
{\small 
\end{document}